\documentclass[showpacs,prl,twocolumn,aps]{revtex4}

\newcommand{\bm}[1]{\mbox{\boldmath$#1$}}
\usepackage{epsfig}\usepackage{psfrag}\usepackage{bm}
\begin{document}
\title{Vortex states of rapidly rotating dilute Bose-Einstein condensates}
\author{Uwe R. Fischer and Gordon Baym}
\affiliation{Department of Physics, University of Illinois 
at Urbana-Champaign, 1110 W. Green Street, Urbana, IL 61801-3080}
\date{\today}

\begin{abstract}

    We show that, in the Thomas-Fermi regime, 
the cores of vortices in rotating dilute Bose-Einstein condensates
adjust in radius as the rotation velocity, $\Omega$, grows, thus precluding a
phase transition associated with core overlap at high vortex density.  In both
a harmonic trap and a rotating hard-walled bucket, the core size approaches a
limiting fraction of the intervortex spacing.  At large rotation speeds, a
system confined in a bucket develops, within Thomas-Fermi, a
hole along the rotation axis, and eventually makes a transition to a giant
vortex state with all the vorticity contained in the hole.

\end{abstract}



\maketitle


    The energetically favored state of rotating superfluids such as $^4$He
\cite{Yarmchuk} and the alkali gas Bose-Einstein condensates
\cite{madison,VortexLatticeBEC,VortexRaman} is a triangular lattice of singly
quantized vortices \cite{Tkachenko66}.  Here we pose the question of the
structure of the lattice and ultimate fate of the superfluid at rotation
speeds so large that a sizable fraction of the fluid is filled by the vortex
cores.  A Type II superconductor with a vortex array simply becomes {\em
normal} at a high magnetic field, $H_{c2}$, where the vortex cores begin to
overlap \cite{abrikosov}.  While in a rotating Bose superfluid the cores of
quantized vortex lines (if of fixed size) would completely fill the system at
an upper critical rotation speed $\Omega_{c2}$ analogous to $H_{c2}$, a low
temperature bulk 
bosonic system does not have a simple normal phase to which it can
return, and thus the problem \cite{Fetter2001}.  
In He II the critical rotation rates at which
the vortex cores approach each other are unobservably large, $\Omega_{c2} \sim
10^{12}$ rad/sec.  In Bose condensed dilute atomic gases, by contrast,
inertial forces can be comparable to interatomic forces, and the approach to
tightly packed vortex lattices is within reach of experiment; indeed current
experiments in harmonically trapped gases achieve rotational velocities
$\Omega$ that are a significant fraction of $\Omega_{c2}$
\cite{madison,VortexLatticeBEC,VortexRaman,HaljanCornell,Rosenbusch}.

We consider a Bose gas at zero temperature in the Thomas-Fermi regime, 
with weak repulsive
interactions described by an s-wave scattering length $a_s$, and  
integrate out (in the sense of the renormalization group) the short range
structure on scales of the vortex separation. 
The structure of the cores and the destiny of the vortex
lattice at high rotational speeds depends fundamentally on the geometry of the
container confining the fluid.  We study here the modification of the lattice
induced by high rotational speeds in both harmonic traps and in cylindrical
hard-walled (square-well) buckets \cite{Kuga}.  In a harmonic trap, where
considerable work has been done on understanding the states of the fluid
\cite{Kavoulakis,Fetter2001}, the radial trapping potential, $V_\perp =
\frac12 m \omega_\perp^2 r^2$, 
where $r$ is the perpendicular distance from 
the axis of rotation, dominates the centrifugal potential.
As the rotational frequency $\Omega$ approaches $\omega_\perp$, the system 
becomes quasi-two dimensional and should eventually enter
a quantum Hall-like state
\cite{Cooper,Viefers,Ho,SHM}; we do not treat this limit here.
We find that in a trap the cross-sectional area
occupied by the vortex cores grows until they fill a limiting fraction, $\sim
1/2$, of the space of the system, and they never touch, even for rotational
velocities arbitrarily close to $\omega_\perp$.  This result goes beyond that
of \cite{Fetter2001}, in that we take into account (albeit approximately) the
modification of the vortex core in the finite sized cell of the lattice, for
given particle number in the cell.

    In a cylindrical bucket, on the other hand, where the centrifugal force
tends to throw the fluid against the walls, we find that at a critical
rotation speed, $\Omega_h = \sqrt 2 \hbar /m\xi_0 R$, the fluid begins to
develop a cylindrical hole in the center which grows with increasing $\Omega$;
here $R$ is the bucket radius, $\xi_0 = \hbar(2mg\bar n)^{-1/2}$ is the
coherence or healing length, $g=4\pi\hbar^2a_s/m$, and $\bar n$ is the uniform
density in a non-rotating cloud.  In the bucket, the ratio of cross-sectional
core area to the
area of the unit cell of the (2D) vortex lattice is independent of $\Omega$
for $\Omega>\Omega_h$, and given by $9\xi_0/\sqrt 2 R$.  A phase transition
associated with core overlap at high vortex density is thus precluded for
either trapping geometry.  In addition, as we show, a system in a bucket
eventually undergoes a phase transition in the limit of large $\Omega$ to a
giant vortex phase with the vorticity concentrated along the cylinder axis.

    The scale of rotation velocities in a dilute gas in a container is set by
$\Omega_0 = 2g\bar n/\hbar = \hbar/m\xi_0^2 \sim \Omega_{c2}$.  For $\xi_0 \simeq 0.2\mu$m
\cite{madison,VortexLatticeBEC}, $\Omega_0\simeq 1.8\times 10^4$ rad/sec in
$^{87}$Rb and $\simeq 6.9\times 10^4$ rad/sec in $^{23}$Na.

    We describe the ground state of the system with a rotating vortex lattice
by an order parameter, $\psi({\bm r})$, determined by minimizing
the energy, $E' = E- \Omega L_z$, in the rotating frame by means of a
variational calculation; here $L_z$ is the component of the angular momentum
of the system along the rotation axis.  With $\hbar=1$,
\begin{eqnarray}
 E'& = &
  \int d^3 r \left[\frac{1}{2m} \left|\left(
   -i\nabla  -m {\bm \Omega} \times {\bm r}\right)\psi\right|^2 \right.
     \nonumber\\
   & & \qquad\quad\left.
   +\left(V(\bm r)-\frac12 m \Omega^2 r^2\right) |\psi|^2 + \frac12 g
|\psi|^4\right]\!\!,
\label{Hprimed}
\end{eqnarray}
where $V(\bm r)$ is the trapping potential.  For a single vortex line of
unit winding number on the cylindrical axis, $\psi({\bm r} ) =
f(r,z)e^{i\phi}$ in polar coordinates.  To calculate $E'$ we employ a
Wigner-Seitz approximation \cite{WS}, replacing the triangular shaped cells of
the (2D) lattice by cylindrical cells of equal radius $\ell$; the assumption
of equal cell area is consistent with observed lattices in traps
\cite{VortexLatticeBEC}.  We assume straight vortex lines for simplicity, and
that the lattice rotates at a uniform angular velocity $\Omega_v$, related to
the (2D) density of vortices $n_v = 1/\pi\ell^2$ by $\Omega_v = \pi n_v/m =
1/m\ell^2$; $\Omega_v$ is in general smaller than the stirring velocity
$\Omega$ by terms of order $1/mR^2$.

    In general, $\psi$ in a cell is a unique function of the number of
particles therein, and can be constructed, e.g., by the method of matched
asymptotic expansion \cite{asymp}.  It is sufficient here to use the simple
approximation that within a cell, labelled by $i$,
\renewcommand{\arraystretch}{1.5}\begin{eqnarray}
 f(x,z) & = &\left\{\begin{array}{l} (x/\xi)\sqrt{n_i(z)},
 \qquad 0 \le x \le \xi,  \\
 \sqrt{n_i(z)}, \qquad  \xi \le x \le \ell,
\end{array}\right.
\label{fAnsatz}\renewcommand{\arraystretch}{1.0}
\end{eqnarray}
where $x$ is the (polar) radial coordinate measured from the center ${\bm
r}_i$ of the cell.  To account for the density variation in the system we
allow the mean density $n({\bm r}_i,z)$ within a cell centered on ${\bm r}_i$
to depend on the radial position of the cell and height.  Ansatz
(\ref{fAnsatz}) is consistent with the exact boundary conditions,
$f(x\to0,z)\sim
x$, and $df(x,z)/dx|_{x=\ell} =0$.  While the core radius in a cell should
depend on the local density, for simplicity we take $\xi$ to depend only on
the mean density in the system, an approximation adequate to bring out the
essential physics of rotation at high speeds in the Thomas-Fermi regime.
As one approaches the mean-field quantum Hall regime \cite{Ho}, it
becomes necessary to include possible dependence of $f$ on the trapping
frequency as well.

    The mean density $n({\bm r}_i,z)$ in the cell is
\begin{eqnarray}
  n({\bm r}_i,z) = \frac{1}{\pi \ell^2}\int_0^{\ell} d^2{r}\,|\psi|^2
  = n_i(z)(1- \zeta/2),
\label{nbarn0}
\end{eqnarray}
where $\zeta \equiv (\xi/\ell)^2 \le 1$ is the ratio of the core area to
the transverse area of the unit cell of the lattice.

    Writing the local velocity as ${\bm v} = \Omega_v r \,\hat {\bm \phi} +
\delta {\bm v}$ to evaluate the kinetic energy in each cell, as in
\cite{baymchandler}, we find
\begin{eqnarray}
 E  & = &   \sum_i E_i
 \to \int d^3{r}\, n({\bm r}) \left(\frac12 m\Omega_v^2 r^2 +(a +
1)\Omega_v
   \right. \nonumber \\
 & & \hspace*{8em}
 +V(\bm r)  + \frac12 gn({\bm r})b\left.\right),
 \label{Elab}
\end{eqnarray}
where we replace the sum over cells by a continuous integration, with
${\bm r}_i \to {\bm r}$.  In Eq.~(\ref{Elab}) the kinetic energy per cell is
$\pi a(\zeta) n({\bm r})/m$ and the interaction energy per cell is $gn^2({\bm
r}) b(\zeta) \pi \ell^2 /2$, where
\begin{eqnarray}
 a &=&\frac{1}{ n({\bm r})}
 \int_0^\ell d {x} x \left[
 \left(\frac{f}{x}\right)^2 + {\left(\frac{\partial f}{\partial
  x}\right )^2}\right]
   = \frac{1-\ln\sqrt{\zeta}}{1- \zeta/2},
    \nonumber\\
 b & = & \frac{2}{\ell^2 n^2({\bm r})}
 \int_0^\ell d{x}x  f^4(x,z)
 = \frac{1-2\zeta/3}{(1- \zeta/2)^2};
 \label{abdef}
\end{eqnarray}
$a(\zeta)$ has a minimum at $\zeta \simeq 0.56$, while $b(\zeta)$
increases slowly with $\zeta$.  The first term in (\ref{Elab}) is the energy
of solid body rotation of the lattice, while the term $\sim n({\bf r})
\Omega_v$ arises from $\Omega_v r \,\hat {\bm \phi}\cdot \delta {\bm v}$.  The
result (\ref{Elab}) is effectively the Thomas-Fermi approximation for the
coarse grained structure, described by $n({\bm r})$, with the functions $a$
and $b$ accounting for the fine grained vortex structure.

    The total angular momentum in the Wigner-Seitz approximation is
\cite{baymchandler} \begin{equation} L_z = \int d^3{r}\, n({\bm r}) \left(1 +
\Omega_v mr^2 \right) \end{equation} the sum of the angular momentum per
vortex cell and the solid body rotation contribution; $L_z$ is independent of
$a$ and $b$.  {\em In toto},
\begin{eqnarray}
 E' &=&  \int d^3{r}\, n({\bm r})
  \left[\left(\frac{\Omega_v^2}{2} -\Omega_v \Omega\right)mr^2
     +(a+1)\Omega_v -\Omega  \right.\nonumber\\
  & & \hspace*{6em}
\left. +V(\bm r) + \frac12 gn({\bm r})b\right].
\label{ErotOmegav}
\end{eqnarray}
Minimization of Eq.~(\ref{ErotOmegav}) with respect to $\Omega_v$ implies
that $\Omega_v = \Omega - (a+1)N/I$, where $I=\int d^3{r}\, mr^2 n(\bm r)$ is
the moment of inertia.  For large systems, $mR^2\Omega\gg1$, we neglect
the small difference of $\Omega_v$ and $\Omega$; then
\begin{eqnarray}
 E' &=&  \int d^3{r}\, n({\bm r})
  \left[-\frac{\Omega^2}{2}mr^2 + V(\bm r)
+  a\Omega + \frac12 gn({\bm r})b\right].\nonumber\\
\label{Erot}
\end{eqnarray}
Minimization of Eq.~(\ref{Erot}) with
respect to $n({\bm r})$ at fixed total particle number yields the
Thomas-Fermi result,
\begin{eqnarray}
 n(\bm r)
  = [\tilde\mu+ m\Omega^2r^2/2 - V(\bm r)]/gb,
  \label{n}
\end{eqnarray}
where $\mu$ is the chemical potential and $\tilde\mu = \mu - a\Omega$.  In
a trap with transverse frequency $\omega_\perp$ and longitudinal frequency
$\omega_z$, the density has the usual form \cite{baympethickPRL96}
\begin{eqnarray}
   n_{\rm trap}(\bm r) = (\tilde\mu/gb)\left(1- (r/R_t)^2 - (z/Z_t)^2\right),
\end{eqnarray}
with $R_t^2=(2\tilde\mu/m)/(\omega_\perp^2 - \Omega^2)$ and
$Z_t^2=(2\tilde\mu/m)/\omega_z^2$.  Then $N=8\pi Z_tR_t^2\tilde\mu/15gb$ and
\begin{eqnarray}
  \frac{\tilde \mu}{\omega_\perp} =
  \frac12\left[15 Nb a_s \sqrt{m\omega_\perp}
   \frac{\omega_z}{\omega_\perp}
      \left(1-\frac{\Omega^2}{\omega_\perp^2}\right)\right]^{2/5}.
  \label{mutilde}
\end{eqnarray}

    By contrast, the density in a cylinder grows quadratically with $r$, until
at a critical velocity $\Omega_h$ where $\tilde \mu =0$, the system develops a
cylindrical hole in the center; in a trap the quadratic trapping potential
dominates, preventing formation of a hole.  Integrating Eq.  (\ref{n}) over
the volume of the cylinder, we find that the hole begins to develop at
$\Omega_h = \sqrt{2b}\Omega_0\xi_0/R = (2b\Omega_0/mR^2)^{1/2}$.  Then
\renewcommand{\arraystretch}{1.75}
\begin{eqnarray}
 n_{\rm cyl}(r) = \left\{\begin{array}{lcr}
\bar n + m\Omega^2\left(
r^2-R^2/2 \right)/2gb, \quad
  \Omega < \Omega_h,\\
{m\Omega^2}(r^2-R_h^2)/2gb, \quad
\Omega\ge  \Omega_h ;
\end{array}\right.
\label{HoleDensity}
\end{eqnarray}
where the radius of the hole is $R_h=R\sqrt{1-\Omega_h/\Omega}$.  The
phase of the order parameter winds by $2\pi m\Omega R_h^2$ at the inner edge
of the hole, corresponding to an array of phantom vortices of density $n_v$
inside the hole.

    The energy per particle in the rotating frame is
\begin{eqnarray}
 \frac{E'_{\rm trap}}{N} & = & \frac{5\tilde\mu}{7}+a\Omega , \\
  \frac{E'_{\rm cyl}}{N} & = &
 -\frac{mR^2\Omega^2}{4}
  +a\Omega
  +\frac{\Omega_0 b}{4}
  - \frac{b\Omega_0}{12}
  \left(\frac{\Omega}{\Omega_h}\right)^4\!, \Omega < \Omega_h
   \nonumber\\
\frac{E'_{\rm cyl}}{N} & = &
 -\frac{m R^2 \Omega^2}{2} +a\Omega  + \frac{m
   \Omega R^2\Omega_h}3,\; \Omega \ge \Omega_h.
  \label{EnergyRotFrame}\renewcommand{\arraystretch}{1.0}
\end{eqnarray}

\begin{center}
\begin{figure}[hbt]
\psfrag{Omega}{\large ${\Omega}/\omega_\perp$}\psfrag{(a)}{\large (a)}
\psfrag{(b)}{\,\,\,\,\,\large (b)}\psfrag{Omegacyl}{\large ${\Omega}/\Omega_0$}
\psfrag{ZetaTrap}{\large $\zeta_{\rm trap} $}\psfrag{zetacyl}{\large $\zeta_{\rm cyl} $}
\psfrag{ZetaMax}{$\zeta_{\rm max}=0.56$}\psfrag{omegaperp}{$=\omega_\perp$}
\psfrag{smallOmega}{\small ${\Omega}/{\Omega_0}$}\psfrag{xi}{\small $\xi/\xi_0$}
\psfrag{Omega Increases}{\footnotesize $\Omega$ increases}\psfrag{Arrow}{$\Longrightarrow$}
\epsfxsize=0.44\textwidth\epsfbox{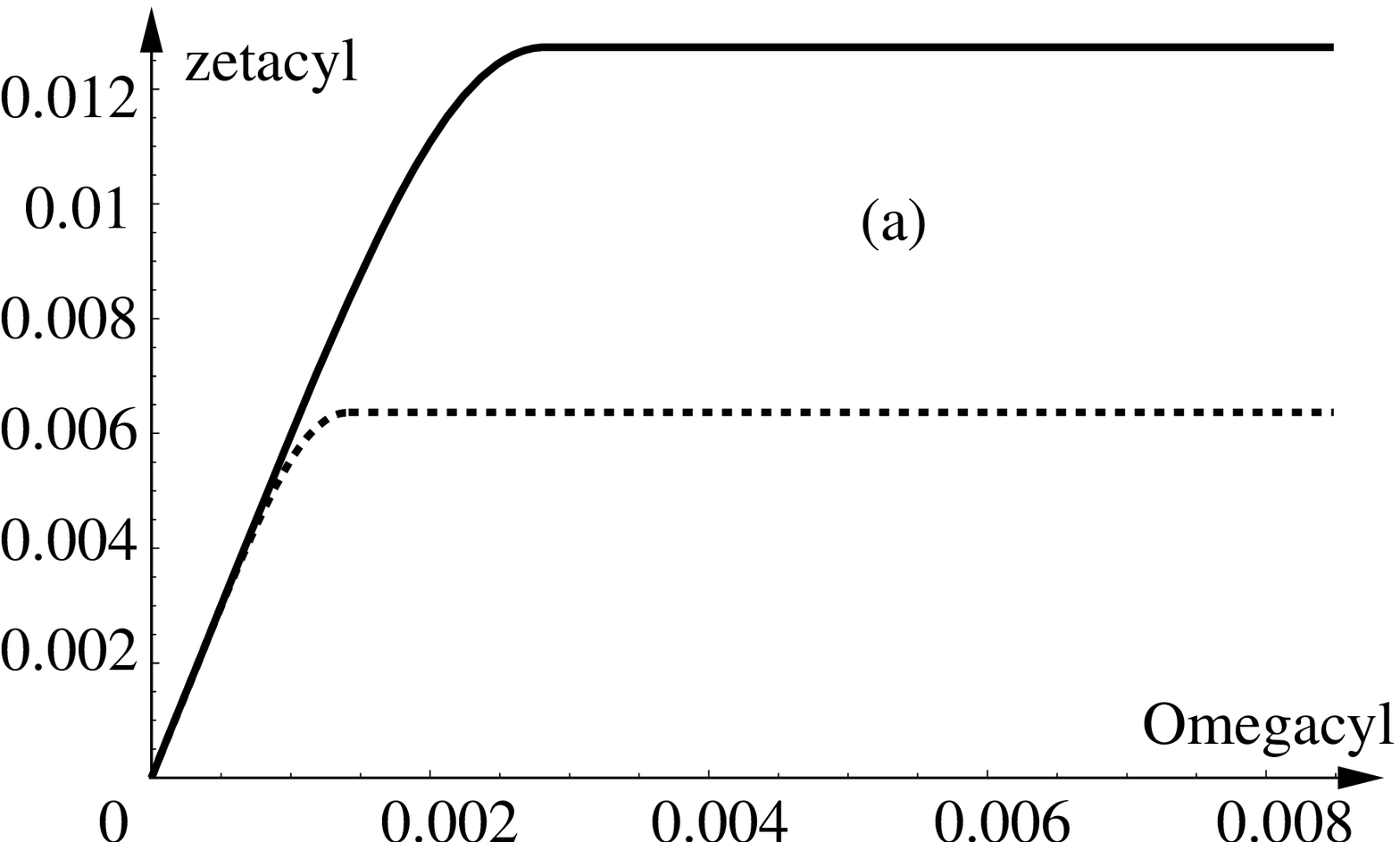}\vspace*{1.5em}
\epsfxsize=0.41\textwidth\epsfbox{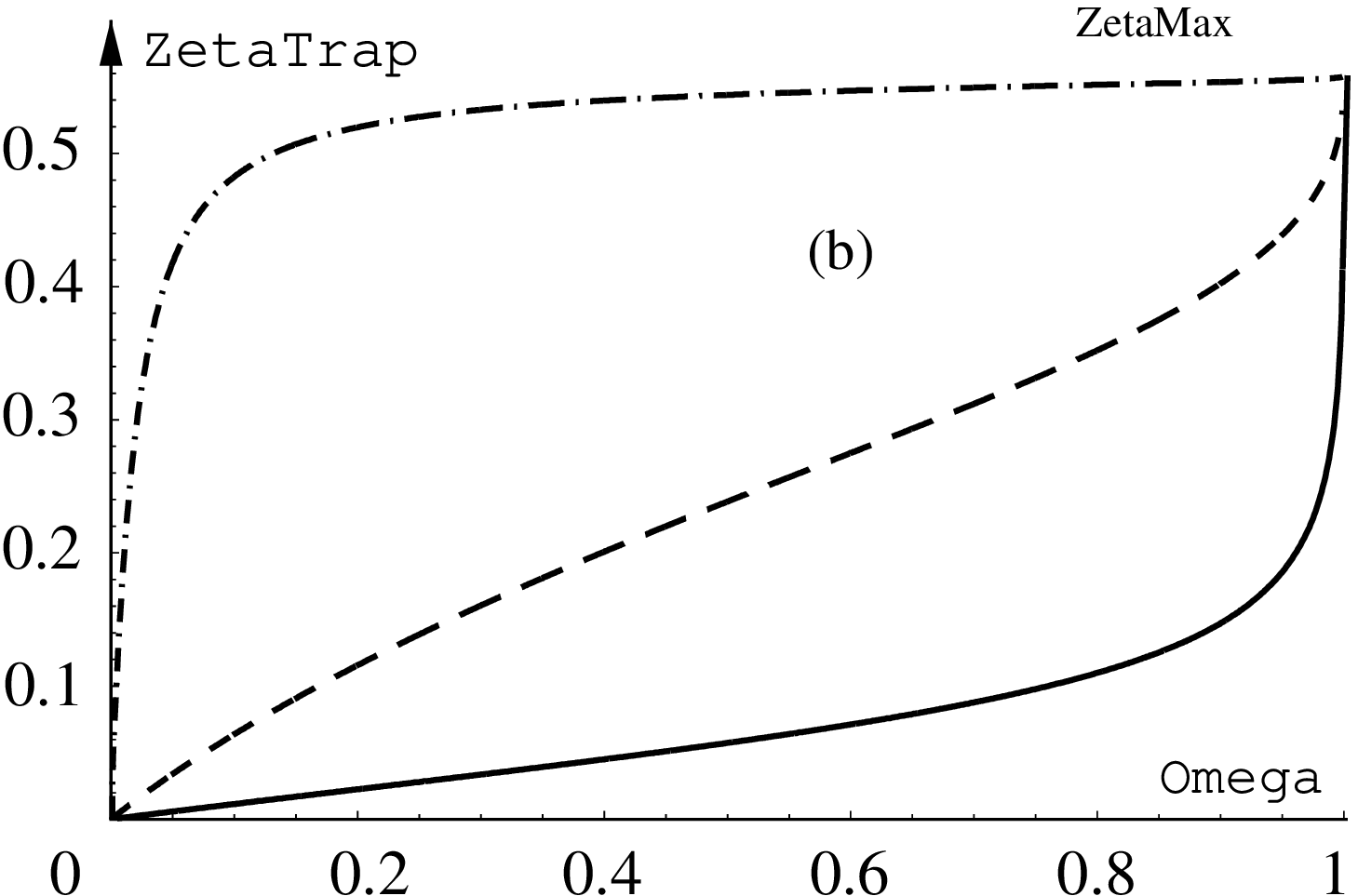}
\caption{\label{Fig1}
Vortex core area as a fraction of the Wigner-Seitz cell area:  (a) $\zeta$
vs.  ${\Omega}/\Omega_0$ in a hard-walled bucket with $R/\xi_0=500$ (solid)
and $R/\xi_0 =1000$ (dashed); (b) $\zeta$ vs.  ${\Omega}/\omega_\perp$ in an
anisotropic cylindrically symmetric harmonic trap with parameters
$N=5\times10^7$, $a_s/d_\perp = 0.01$, and $\omega_\perp/\omega_z = 10^2$
(solid line), $10^4$ (dashed), and $10^8$ (dash-dotted).}
\end{figure}
\end{center}

 The core size is determined by minimizing $E'$ with respect to $\zeta$ at
fixed $N$ and $\Omega$; differentiating Eq.~(\ref{Erot}) directly we have
\begin{equation}
 \Omega\frac{\partial a}{\partial \zeta} + \frac{g \langle n \rangle}{2}
    \frac{\partial b}{\partial \zeta}= 0
\label{meann}
\end{equation}
where $\langle n\rangle = \int n^2(r) /\int n(r)$ is a mean of the density
in the system.  [Note that were we, more correctly, to solve for the vortex
structure within each cell, then the integration in $\langle n \rangle$ in
Eq.~(\ref{meann}) would be only over the given cell.] In a bucket the relevant
regime is small $\zeta$, where $b(\zeta)\simeq 1+\zeta/3$, and $\partial
a/\partial \zeta \simeq -1/2\zeta$, the minimum is thus at $\zeta_{\rm cyl}
=3\Omega/g\langle n\rangle$.  For $\Omega < \Omega_h$, $\langle n \rangle =
{\bar n} \left(1+ \frac13(\Omega/\Omega_h)^4\right)$, and
\begin{equation}
\zeta_{\rm cyl}
\simeq \frac{6\Omega/\Omega_0}{1+\frac13(\Omega/\Omega_h)^4},
\qquad \Omega < \Omega_h.
\end{equation}
For $\Omega \ge \Omega_h$, $\langle n \rangle = \frac43 {\bar
n}\Omega/\Omega_h$, and
\begin{equation}
  \zeta_{\rm cyl} = \frac92 \frac{\Omega_h}{\Omega_0}, \qquad \Omega
\ge \Omega_h.
  \label{zetacyl}
\end{equation}
The total area occupied by the vortex cores thus scales as the ratio of
zero temperature coherence length and system size and is independent of
$\Omega$, see Fig.~\ref{Fig1}a.

    In a trap, $\langle n \rangle = 4\tilde \mu/7gb$; for small $\Omega$,
$\zeta \ll 1$ and $\partial a/\partial \zeta \simeq -1/2\zeta$, so that $\xi$
grows with increasing $\Omega$ as $(\omega_\perp^2-\Omega^2)^{-1/5}$, as in
\cite{Fetter2001}.  However, as $\Omega\to \omega_\perp$, $\tilde\mu$ and
therefore $\langle n \rangle \to 0$, and the solution of (\ref{meann})
requires $\partial a/ \partial \zeta \to 0$, which occurs at $\zeta\simeq
1/2$; this is the limiting value of $\zeta$.  Quite generally, the fact that
$\langle n \rangle \to 0$ as $\Omega\to \omega_\perp$ implies that the
solution must minimize the kinetic energy in Eq.~(\ref{Hprimed}).  The cores
can never overlap, and the system can, within mean-field theory,
maintain a lattice of quantized vortices
for all $\Omega < \omega_\perp$.  This behavior is shown in Fig.~\ref{Fig1}b;
in strongly elongated traps, as in a cylinder, the system reaches a {\em
self-similar} regime in which the core size scales with the intervortex
spacing.

    The rotation speeds needed to observe the increase of core size with
$\Omega$ are in principle accessible within current harmonic trap experiments, 
prior to the onset of a quantum Hall state.  In \cite{VortexLatticeBEC}, $N
\sim 5\times 10^7$, $\omega_\perp/\omega_z =4.2$,
$\Omega/\omega_\perp\lesssim$ 0.7, and $\omega_\perp/\Omega_0=0.008$; an
increase in aspect ratio by a factor 24, corresponding to the solid curve in
Fig.  \ref{Fig1}b, should be sufficient to observe the strong increase of the
core size for $\Omega \lesssim \omega_\perp$.

At sufficiently large $\Omega$, a dilute gas in a cylindrical trap makes a
transition to a giant vortex state, with vorticity concentrated in the center
rather than spread throughout the system.  Such behavior is seen in numerical
simulations in traps that rise faster than harmonic \cite{lundh,tku} (and also
in the presence of vortex pinning \cite{simula}). Dynamical
formation of metastable giant vortices in harmonic traps with 
7 to 60 units of vorticity is reported in Ref.  \cite{JILAgiant}. 
 At the level of
Thomas-Fermi, which is adequate to describe the giant vortex regime in large
systems ($mR^2\Omega\gg1$) a giant vortex, described by order parameter,
$\psi({\bm r}) = \sqrt{n_{\rm G}(r)} e^{i\nu\phi}$; has an energy in the rotating
frame,
\begin{eqnarray}
{E_{\rm G}'}
  = \int d^3{r}\,  n_{\rm G}({\bm r}) \left(\frac{\nu^2}{2mr^2}
     + \frac g2  n_{\rm G}({\bm r})  - \nu {\Omega}\right).
\label{EGiantnu}
\end{eqnarray}
Minimizing with respect to $n({\bm r})$ we find the analog of
Eq.~(\ref{n}), $n_{\rm G}(\bm r) = (\tilde\mu - \nu^2/2 mr^2)/g$,
where $\tilde\mu = \mu + \nu\Omega$.  
In Thomas-Fermi, the giant vortex has a
hole in the center of radius $R_{\rm G}= \alpha R$, where $\alpha =
\nu/R\sqrt{2m\tilde\mu}$.  
In the limit of large $\Omega$ ($\gg \Omega_h$),
where
$\alpha\to
1$, we expand in powers of $(\Omega_0/mR^2\Omega^2)^{1/2}$ and find
\begin{equation}
 \frac{E'_{\rm G}}{N} =
   -\frac12 m R^2 \Omega^2 + \frac{\sqrt{2mR^2\Omega_0}}3 \Omega
     +\frac{\Omega_0}{18} + \cdots .
\label{EnegyGiant}
\end{equation}
Comparing with the lattice energy, Eq.~(\ref{EnergyRotFrame}), and using
(\ref{zetacyl}), we find the transition to the giant vortex state at
\begin{equation}
  \Omega_{\rm G} =  \frac{\Omega_0}{9(1+2 a(9\xi/\sqrt2 R))}
  \simeq \frac{\Omega_0}{9(1+\ln(R/\xi_0))}.
\end{equation}

    In this paper we have studied only the zero temperature structure of
rotating Bose gases in the long wavelength Thomas-Fermi regime.  
The full structure at finite temperatures reflected in
the phase diagram in the $\Omega$--$T$ plane (studied for a harmonically
trapped gas in \cite{StringariPRL99}) may be rich.  For example, at
sufficiently large rotation velocities and low temperatures such that thermally
excited motion perpendicular to the boundary becomes frozen out, the fluid
forced against the boundary of the bucket becomes effectively two-dimensional.
Under these circumstances the fluid can possibly undergo a transition to a
two-dimensional Kosterlitz-Thouless phase.

    Author G.\,B. is grateful to Chris Pethick for discussions that formulated
this approach to the present problem, and to Jason Ho for pointing out the
possible development of a Kosterlitz-Thouless state at high rotational
velocities.  Author U.\,R.\,F. acknowledges financial support by the {\it Deutsche
Forschungsgemeinschaft} (FI 690/2-1).  This research was supported in part by
NSF Grants PHY98-00978, PHY00-98353, and DMR99-86199.


\begin{thebibliography}{499}

    \bibitem{Yarmchuk} E.\,J. Yarmchuk, M.\,J.\,V. Gordon, and R.\,E. Packard,
Phys.  Rev.  Lett.  {\bf 43}, 214 (1979).

    \bibitem{madison} K.\,W. Madison, F. Chevy, W. Wohlleben, and J. Dalibard,
Phys.  Rev.  Lett.  {\bf 84}, 806 (2000); F. Chevy, K.\,W. Madison, and J.
Dalibard, Phys.  Rev.  Lett.  {\bf 85}, 2223 (2000); K.\,W. Madison, F. Chevy,
V. Bretin, and J. Dalibard, Phys.  Rev.  Lett.  {\bf 86}, 4443 (2001).

    \bibitem{VortexLatticeBEC} J.\,R.  Abo-Shaeer, 
C. Raman, J.\,M.  Vogels, and
W. Ketterle, Science {\bf 292}, 476 (2001).  This paper clearly depicts vortex
cores filling a fraction of the sample area in a dilute alkali gas 
Bose-Einstein condensate.

    \bibitem{VortexRaman} C. Raman, J.\,R. Abo-Shaeer, J.\,M. Vogels, 
K. Xu, and W. Ketterle, Phys. Rev. Lett. {\bf 87}, 210402 (2001).

    \bibitem{Tkachenko66} V.\,K. Tkachenko, Zh.  \'Eksp.  Teor.  Fiz.  {\bf 49},
1875 (1965) [Sov.  Phys.  JETP {\bf 22}, 1282 (1966)]; Zh.  \'Eksp.  Teor.  Fiz.
{\bf 50}, 1573 (1966) [Sov.  Phys.  JETP {\bf 23}, 1049 (1966)].

    \bibitem{abrikosov} A.\,A. Abrikosov, Zh.  \'Eksp.  Teor.  Fiz.  {\bf 32},
1442 (1957) [Sov.  Phys.  JETP {\bf 5}, 1174 (1957)].

\bibitem{Fetter2001} A.\,L. Fetter, Phys. Rev. A {\bf 64}, 063608 (2001).

    \bibitem{HaljanCornell} P.\,C.  Haljan, I. Coddington, P. Engels, 
and E.\,A.Cornell, Phys.  Rev.  Lett.  {\bf 87}, 210403 (2001), 
and P. Engels, I. Coddington, P.\,C.  Haljan, and E.\,A.  Cornell, 
Phys. Rev. Lett. {\bf 89}, 100403 (2002)   
infer rotation
rates of up to 95\% of the centrifugal limit of the transverse trap frequency.
The latter reports evidence for a new phase in
highly distorted lattices in which the vortex cores appear to merge into
sheets.  Whether such a phase can exist is not clear.  Explanations in terms
of dynamics of discrete vortices have been given by 
A.\,A. Penckwitt, R.\,J. Ballagh, and C.\,W. Gardiner,
Phys. Rev. Lett. {\bf 89}, 260402 (2002), and E.\,J.  Mueller and T.-L.
Ho, cond-mat/0210276.  Furthermore, a sheet singularity is not permitted by the
(single component) Gross-Pitaevski\v\i\/ equation (L.\,P. Pitaevski\v\i\/, private communication to GB).

    \bibitem{Rosenbusch} P. Rosenbusch, D.\,S.  Petrov, S. Sinha, F. Chevy, V.
Bretin, Y. Castin, G. Shlyapnikov, and J. Dalibard, Phys. Rev. Lett. {\bf 88},
250403 (2002).
                 

    \bibitem{Kuga} We study the cylindrical bucket as an extreme tractable
case of a radial potential steeper than harmonic.  Such potentials may be
created using Laguerre-Gaussian (doughnut) laser beams; T. Kuga {et al.},
Phys.  Rev.  Lett.  {\bf 78}, 4713 (1997).

    \bibitem{Kavoulakis} A.\,D.  Jackson and G.\,M.  Kavoulakis, Phys.  Rev.
Lett.  {\bf 85}, 2854 (2000); G.\,M.  Kavoulakis, B. Mottelson, and C.\,J.
Pethick, Phys.  Rev.  A {\bf 62}, 063605 (2000); A.\,D.  Jackson, G.\,M.
Kavoulakis, B. Mottelson, and S.\,M.  Reimann, Phys.  Rev.  Lett.  {\bf 86}, 945
(2001).

   

    \bibitem{Cooper} N.\,K.  Wilkin, J.\,M.\,F.  Gunn, and R.\,A.  Smith, Phys.  Rev.
Lett.  {\bf 80}, 2265 (1998); N.\,K.  Wilkin and J.\,M.\,F.  Gunn, Phys.  Rev.
Lett.  {\bf 84}, 6 (2000); N.\,R.  Cooper, N.\,K.  Wilkin, and J.\,M.\,F.  
Gunn, Phys. Rev.  Lett.  {\bf 87}, 120405 (2001).

    \bibitem{Viefers} S. Viefers, T.\,H. Hansson, and S.\,M. Reimann, Phys.
Rev.  A {\bf 62}, 053604 (2000).

    \bibitem{Ho} T.-L.  Ho, Phys.  Rev.  Lett.  {\bf 87}, 060403 (2001).

\bibitem{SHM} J. Sinova, C.\,B.  Hanna, and A.\,H.  MacDonald, Phys.  Rev.
Lett.  {\bf 89}, 030403 (2002).

    \bibitem{WS} E. Wigner and F. Seitz, Phys.  Rev.  {\bf 43}, 804 (1933);
Phys. Rev. {\bf 46}, 509 (1934).

    \bibitem{asymp} B.\,Y.  Rubinstein and L.\,M.  
Pismen, Physica D {\bf 78}, 1
(1994); J.\,R. Anglin, Phys. Rev. A {\bf 65}, 063611 (2002); 
A.\,L. Fetter and A.\,A. Svidzinsky,  J. Phys.: Condens. Matter {\bf 13}, R135 (2001).

    \bibitem{baymchandler} G. Baym and E. Chandler, J. Low Temp.  Phys.  {\bf
50}, 57 (1983); J. Low Temp.  Phys. {\bf 62}, 119 (1986).

    \bibitem{baympethickPRL96} G. Baym and C.\,J.  Pethick, Phys.  Rev.  Lett.
{\bf 76}, 6 (1996).

    \bibitem{lundh} E. Lundh, Phys. Rev. A {\bf 65}, 043604 (2002).

    \bibitem{tku} K. Kasamatsu, M. Tsubota, and M. Ueda, Phys. Rev. A 
{\bf 66}, 053606 (2002).
                         

    \bibitem{simula} T.\,P.  Simula, S.\,M.\,M. Virtanen, and M.\,M. 
Salomaa, Phys.  Rev.  A {\bf 65}, 033614 (2002).

\bibitem{JILAgiant} P. Engels, I. Coddington, P.\,C. Haljan, V.
Schweikhard, and E.\,A. Cornell, cond-mat/0301532.

    \bibitem{StringariPRL99} S. Stringari, Phys.  Rev.  Lett.  {\bf 82}, 4371
(1999).

\end{thebibliography}
\end{document}